\documentclass[a4paper, amsfonts, amssymb, amsmath, reprint, showkeys, nofootinbib, oneside]{revtex4-1}
\usepackage[english]{babel}
\usepackage[utf8]{inputenc}
\usepackage{amsthm}
\usepackage{mathtools}
\usepackage{physics}
\usepackage{xcolor}
\usepackage{graphicx}
\usepackage[left=23mm,right=13mm,top=35mm,columnsep=15pt]{geometry} 
\usepackage{adjustbox}
\usepackage{placeins}
\usepackage[T1]{fontenc}
\usepackage{lipsum}
\usepackage{csquotes}

\usepackage[pdftex, pdftitle={Article}, pdfauthor={Author}]{hyperref} %
\usepackage{array}
\usepackage{longtable}
\usepackage{multirow}

\begin{document}
\title{Design strategies for the self-assembly of polyhedral shells}

\author{Diogo E. P. Pinto$^1$, Petr $\check{\text{S}}$ulc$^{2,3}$, Francesco Sciortino$^1$ and John Russo$^1$}
    \email[Correspondence email address: ]{john.russo@uniroma1.it}%
    \affiliation{$^1$Dipartimento di Fisica, Sapienza Universit\`{a} di Roma, P.le Aldo Moro 5, 00185 Rome, Italy}
    \affiliation{$^2$Life and Medical Sciences Institute (LIMES), University of Bonn, Bonn, Germany}
    \affiliation{$^3$School of Molecular Sciences and Center for Molecular Design and Biomimetics, The Biodesign Institute, Arizona State University, 1001 South McAllister Avenue, Tempe, Arizona 85281, USA}

\date{\today}

\begin{abstract}
The control over the self-assembly of complex structures is a long-standing challenge of material science, especially at the colloidal scale, as the desired assembly pathway is often kinetically derailed by the formation of amorphous aggregates. Here we investigate in detail the problem of the self-assembly of the
three Archimedean shells with five contact points per vertex, 
i.e. the icosahedron, the snub cube, and the snub dodecahedron. We use patchy particles with five interaction sites (or patches) as  model for the building blocks, and recast the assembly problem as a Boolean satisfiability problem (SAT) for the patch-patch interactions. This allows us to find effective designs for all targets, and to selectively suppress unwanted structures.
By tuning the geometrical arrangement and the specific interactions of the patches, we demonstrate 
that lowering the symmetry of the building blocks reduces the number of competing structures, which in turn can considerably increase the yield of the target structure. These results cement SAT-assembly as an invaluable tool to solve inverse design problems.
\end{abstract}

\maketitle

\section{Introduction}

Self-assembly encompasses a large array of phenomena through which materials are formed using simple microscopic building blocks \citep{Whitelam2015}. In nature, many striking examples of self-assembly are found, from virus capsids to lipid bilayers~\citep{Whitesides2002, Parnell2015, Teyssier2015, MartinBravo2021, Hagan2006}, but assembling new synthetic materials has proved to be very challenging. Successful examples of artificial self-assembly have required a large dose of educated guesses~\citep{Dziomkina2005, Glotzer2007, Kim2011, LaCour2022,  McGorty2010, Mu2022, Nykypanchuk2008, Sacanna2011, Wang2012, Wang2015, Joshi2016,Bishop2022,Reguera2019, McMullen2022, Wilber2007, Kraft2012, Hatch2016, Halverson2013}. One of the main difficulties resides in how to optimize the geometrical properties or the interactions between the building blocks without leading to competing or kinetically arrested configurations~\citep{Frenkel2011, Lash2015, Blaaderen2006, Meulen2015}.

Here we focus on the self-assembly of finite-size structures and in particular on specific polyhedral shells. From an application standpoint, the potential of closed shells to act as a drug delivery system has been a widely researched topic, where a given drug is encapsulated within a closed shell and then driven to a specific diseased area where the drug is locally released such that the least amount of non diseased tissue is affected \citep{Huang2007, Uchida2007}. For this, the shell needs to close around a specific reagent and then open when external conditions are met. Recently, there have been suitable experimental realizations, for example, using DNA-origami, where selective interactions can be introduced to mimic patchy particles \citep{Mosayebi2017, Lee2022, Jun2021, Rothemund2006, sigl2021programmable}.

When focusing on finite-size shells additional challenges arise compared to the ones encountered in the self-assembly of crystal structures. Firstly, the self-assembly occurs exclusively from the gas phase, which rules out the possibility to use (critical point induced) density fluctuations to accelerate the rate of aggregation~\citep{wolde1997enhancement}, as frequently done in the case of crystals. On the contrary, the formation of finite-size aggregates stabilizes the gas phase to high densities with respect to the liquid phase, possibly introducing a density dependence of the aggregation pathway. Secondly, the small size of the aggregates, compared to the infinitely repeating units of a crystal, can stabilize kinetic traps, i.e. structures whose free-energy is not as low as the one of the target structure, but that require an exceedingly long waiting time to break. Lastly, the formation of finite-size aggregates is a continuous process, and is not accompanied by a phase transition as in crystals. The absence of a critical-size of formation (i.e. the critical nucleus) means that it is not sufficient to suppress a handful of competing structures at one length-scale, but that the assembly process has to proceed without defects at every stage. This problem is reflected in the difficulty of perfectly closing large-size aggregates, such as capsids~\citep{Mosayebi2017}.

Here we show how to successfully tackle these challenges by transforming the self-assembly problem into a Boolean satisfiability problem. This technique, named \emph{SAT-assembly}, was recently introduced by some of us to successfully assemble challenging crystalline structures,
with an emphasis on structures that have photonic applications, such as the tetrastack~\citep{Romano2020a} and diamond cubic crystal~\citep{Romano2020a,Russo2022,Rovigatti2022}, structures with a high number of atoms in the unit cells, such as clathrate crystals~\citep{Romano2020a}. The SAT-assembly design pipeline allows for a fast search of design space for solutions that can form desired target structure and --- equally important --- avoid undesired alternative assemblies or kinetic traps.
In this article we will demonstrate a successful application of the SAT-assembly framework to the assembly of complex polyhedral shells. We focus on the three polyhedra that can assemble from building blocks with a  coordination number of five, i.e. the icosahedron, the snub cube, and the snub dodecahedron.
As building blocks we use patchy particles, which are a model of hard spheres with attractive patches on its surface~\citep{zhang2004self, Bianchi2006, Sciortino2009, Romano2010, Kraft2012, Rovigatti2018, Russo2021}. These represent a coarse-grained approach to describe multiple systems, e.g. colloids, proteins, polymers, DNA origami \citep{pawar2010fabrication, Sacanna2011, Wang2012, ravaine2017synthesis, Rovigatti2022}.
By associating to each patch a colour, we can translate the self-assembly problem into the problem of finding how patch colours should interact to form the target structure, while at the same time avoiding competing structures. It is possible that different natural structures follow similar strategies to the examples explored here, with focus on selectivity. For example, adenoviruses are known for their icosahedral nucleocapsid which is assembled by two main proteins and three minor ones that mostly influence the interactions \citep{Harrison2010}. As such, these processes evolved to require a minimum level of specificity in the building blocks and interactions in order to form regular shells.

In the manuscript we will first show that, except for the smallest structure (the icosahedron), these shells do not assemble correctly if only the geometrical information about the target structure  (the "educated guess") is used. Instead we will introduce multi-coloured designs, i.e. designs where more than one patch type is present. As the number of patch types increases, so does the complexity of the design, making SAT a necessary tool to be able to find the desired interactions between patch colours that can form target shell and avoid alternative assemblies. We will show that the yield of the different structures depends sensitively, in addition to the particular geometrical patch arrangement, on  the number of colours used, but that the different results can be rationalized by looking at the symmetry of the designs: less symmetrical building blocks self-assemble with the highest yield in the target structures. In particular, \emph{chiral} designs, i.e. not having a mirror-plane symmetry, are found to be very effective building blocks regardless of the chirality of the target structure. We also explore the role of geometrical frustration, where the angle between the patches does not match the angle between the particles in the target structure, to create designs where the target structure depends on external conditions (in our case the concentration of building blocks). We also emphasize how SAT is crucial to assemble a given structure while excluding competing ones. We find that mutual exclusion requires complex designs, with multiple patch colours and particle species (different patch colour arrangement in different particles), which lead to higher yields and assemblies that are more robust to geometrical frustrations. Lastly, we discuss how increasing the number of colours and species can slow down the short time kinetics of the assembly process, which emphasizes the design principles discussed here to reach optimal results.

\section{Models}

\begin{figure}[t]
	\includegraphics{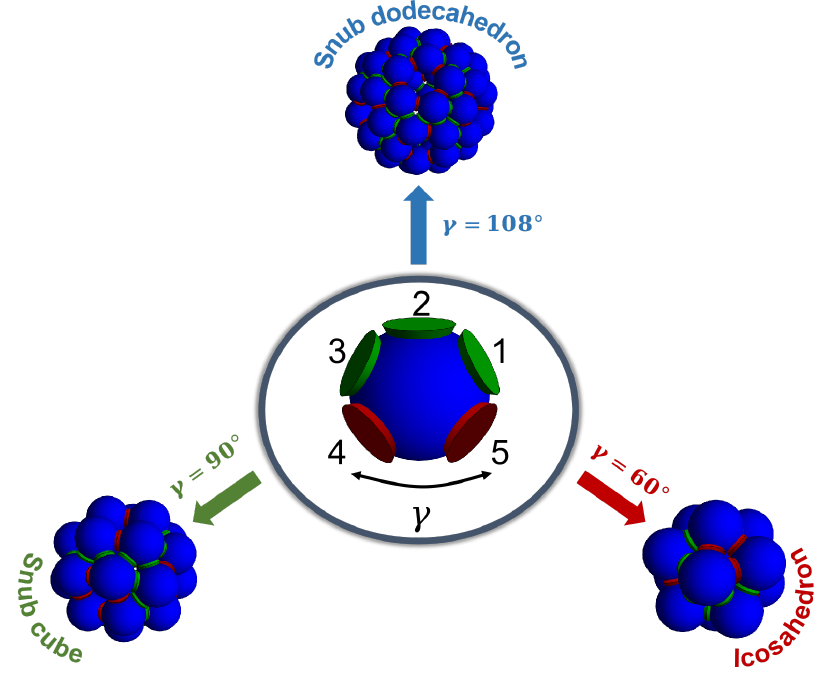}
	\caption{\label{SAT} Schematic representation of the three Archimedean structures with five contact point per vertex and of a valence-five particle with two types of patches (green and red) which according to the SAT algorithm can assemble into the three different polyhedron shells. Note that the green patches only interact with green while the red only interact with red. $\gamma $ indicates the in-plane angle between the two red patches.
	}
\end{figure}

We consider a system composed of $N$ patchy particles in a cubic box of length $L$. The particles are characterized by a hard core of radius $\sigma$ with five patches on its surface located on one side of the particle forming a star-like shape with mirror symmetry along one of the patches (as seen in the center of Fig.~\ref{SAT}).
We follow the same SAT formulation as the one in Ref.~\citep{Russo2022}. We consider that each patch can have a given colour $x_c$ between $1\leq x_c\leq N_c$, where $N_c$ is the total number of distinct colours. These colours can be distributed onto the patches in specific arrangements, each unique sequence can be considered a particle specie $x_p$, thus $1\leq x_p\leq N_p$, where $N_p$ is the total number of distinct species. SAT is then used to find if a given combination of $N_p$ and $N_c$ can satisfy a given polyhedral shell, e.g. if it satisfies all the topological constraints, and a solution/design is calculated which can be used to prepare the composition of the system. In the \emph{Supporting Information} we go into more detail on the different constraints (clauses) used in SAT.

We study the self-assembly outcome of our designs
	focusing on the static yield, defined as the ratio of the number of fully formed target structures over the total number of aggregates. This quantity is measured once the bond probability has reached a (metastable) equilibrium state, that in our case is obtained via Monte Carlo simulations using aggregation-biased moves and using the Kern-Frenkel portential~\citep{Kern2003, bol1982monte} to describe the interaction between patches (See Methods). In Sec.~3\ref{sec:kinetics} we also consider the short-time aggregation kinetics of our designs, i.e. the rate at which particles form aggregates from the solution. To study this we use Molecular Dynamics simulations using a continuous version of the Kern-Frenkel portential (see Methods).

Our self-assembly problem is represented schematically in Fig.~\ref{SAT}. Starting with patchy particles of valence five we wish to selectively assemble three Archimedean polyhedral shells: the 12-particle \emph{icosahedron}, the 24-particle \emph{snub cube}, and the 60-particle \emph{snub dodecahedron}. 
In the ideal structure, three contact points (patches 1, 2 and 3) are in the same location in all three  polyhedral shells, making an in-plane angle with the center of the particle of $60^\circ$, while the other two contact points differ for the in-plane angle that patches 4 and 5 make with the center of the particle, which we call $\gamma$. Note that in the ideal structure $\gamma=60^\circ$ for the icosahedron, $\gamma=90^\circ$ for the snub cube, and  $\gamma=108^\circ$ for the snub dodecahedron. We consider two geometrical control parameters: the angular width of the patch interaction, $\cos\theta_\text{max}$, and the in-plane angle $\gamma$. 
In addition to  the geometrical parameters, we investigate different SAT solutions, i.e. different number of particle types and different colours among the five patches. As alluded before, the SAT-assembly algorithm provides an interaction table among the colours such that all bonds in the target structure(s) are satisfied. Finally we also vary the temperature and density conditions of the assembly to study the phase behaviour of the different designs.

Unless stated otherwise, the temperature used is $T=0.097$. This value is high enough to guarantee that we are above the critical miscelle concentration for the densities explored (see \textit{Supporting Information}). It also guarantees that bonds are able to break at a reasonable pace during the simulations, but still persist long enough for structures to form especially in more dilute systems. For the densities explored, this value also guarantees that we are always close to the ideal gas phase of fully formed aggregates (see \textit{Supporting Information}). Temperature ($T$) is expressed in units of $\varepsilon$, the patches binding energy, and $k_B=1$, while all lengths are in units of $\sigma$.

\section{The "educated guess": Geometrical designs}

\begin{figure}[t]
	\includegraphics{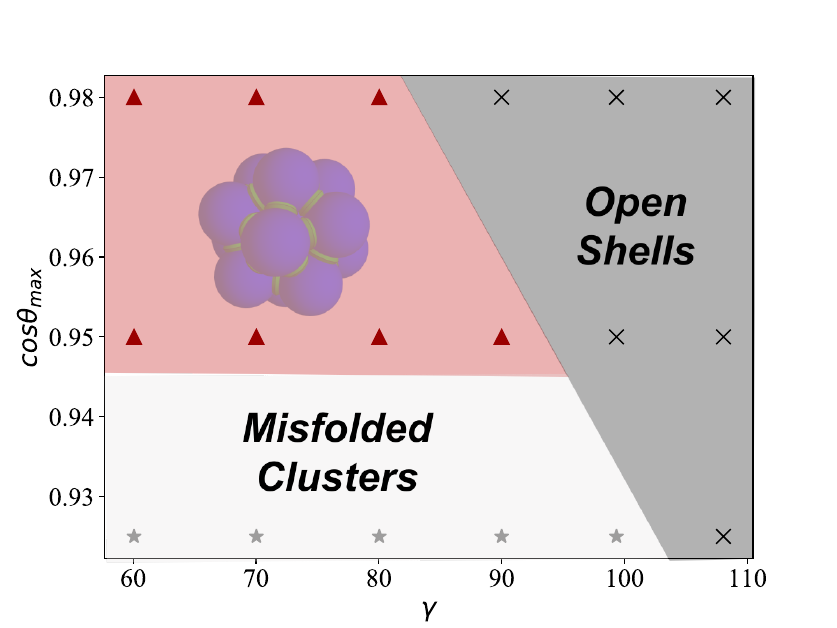}
	\caption{\label{N1c1} Most probable structure formed for different in-plane angles, $\gamma $, and for different patch width, $\cos \theta _{max}$, when all patches are identical. The triangles inside the red region represent parameters where the polyhedron forming with highest yield is the icosahedron. Black crosses represent parameter values where open/incomplete clusters are more probable. The grey stars represent systems where closed clusters are able to form, but none of them is an Archimedean five coordinated polyhedron. Thus, for the one-patch type "educated guess" solution, only the icosahedron is able to form. Results refer to $T=0.097$ and $\rho =0.01$. Temperature ($T$) is expressed in units of $\varepsilon$ and $k_B=1$, while all lengths are in units of $\sigma$.
	}
\end{figure}

To consider geometric effects only, we fix one patch colour (i.e. each patch can interact with all others), and explore different values of the patch width, $\cos\theta_{\rm max}$ as well as different geometries of the patches by changing the in-plane angle $\gamma$.

The former parameter allows for more flexibility of the bonds due to larger patch widths. The latter improves our ability to target the different structures proposed in Fig.~\ref{SAT} by more easily satisfying their geometry.

Figure \ref{N1c1} summarizes the results, showing the most probable closed structure assembled for each value of $\cos\theta_{max}$ and $\gamma$ considered. As represented by the triangle symbols (and the shaded area), we find that only the icosahedron is able to form for the range of parameters explored, while the snub cube and snub dodecahedron fail to assemble even when the geometry of the particles is the ideal one ($\gamma=90^\circ$ for the snub cube, and  $\gamma=108^\circ$ for the snub dodecahedron).

We classify the state points where no target structure is formed into two groups. The grey stars correspond to irregular aggregates, thus clusters more akin to micelles, that close without defects but do not have a regular shape. If no closed structure is formed, we use black crosses to classify the state point, the ones where particles form bonds but they do not close, thus forming an open shell.

These results confirm the well-known fact that self-assembly designs cannot generally rely only on the geometrical properties of the building units alone. In our case, only the smallest target structure, the icosahedron, is successfully self-assembled. The assembly is limited up to $\gamma\sim 90^\circ$, above which the geometry of the building unit is not compatible with bond formation in the target structure and only incomplete structures are formed. As expected, the $\gamma$ range increases for increasing patch width (lowering $\cos\theta_{max}$). But crucially the assembly is also limited at large patch widths, $\cos\theta_{max}>0.95$, below which mostly irregular structures are observed. With large patch widths there are multiple ways for a shell to close onto itself, and this degeneracy entropically stabilizes irregular structures over the more ordered polyhedral shells. We note that although icosahedron structures are still observed for large patch widths, $\cos\theta_{\rm max}<0.95$, a lot fewer are completely assembled (less than 5\%).

\section{SAT designs}

\subsection{Patch colouring}

\begin{figure}[t]
	\includegraphics{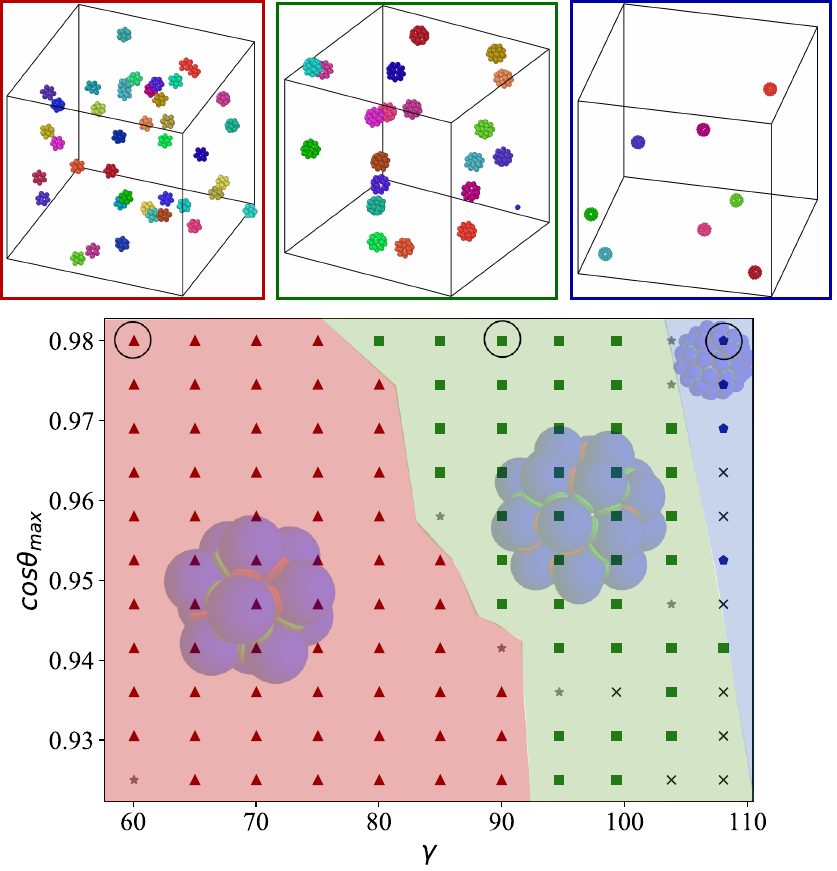}
	\caption{\label{N1c2} Top: Snapshots of the final configurations obtained from simulations based on a one component system. Particles have two types of patches (red and green) and bind via self-complementary interactions (red with red and green with green). In all simulations $\cos \theta_{max}=0.98$ and the $\gamma =60, 90$ and $110$ degrees respectively, i.e. at the optimal value for the three selected structures. The simulation for the icosahedron and snub cube were done with $T=0.097$ and $\rho =0.01$, while for the snub dodecahedron $\rho =0.001$ was used. In all snapshots, the yield of the most probable shell is close to 100\%. Bottom: Most probable structure formed for different in-plane angles, $\gamma $ and for different patch width, $\cos \theta _{max}$. The triangles inside the red region represent parameters where the most probable structure is the icosahedron. The squares inside the green region represent parameters where the most probable structure is the snubcube. The pentagons inside the blue region represent parameters where the most probable structure is the snub dodecahedron. Black crosses represent systems where only open/incomplete clusters are formed. The grey stars represent systems where clusters are able to close but none correspond to the ones in Fig.~\ref{SAT}. Results were calculated for $T=0.097$ and $\rho =0.01$. Temperature ($T$) is expressed in units of $\varepsilon$ and $k_B=1$, while all lengths are in units of $\sigma$.
	}
\end{figure}

In contrast to the previous section, we now break the interaction symmetry, and introduce patch colours.
We start by considering a solution that satisfies all three polyhedral structures with only one particle specie ($N_p=1$) and two patch colours ($N_c=2$). We employ SAT to satisfy such constraints and extract the proper patch ordering and interactions.
The SAT solution, represented schematically  in Fig.~\ref{Yield},
allows only interactions among patches of the same colour (green with green, red with red): $1$, $2$ and $3$ (green) on one side, and $4$ and $5$ (red) on the other.

In Fig.~\ref{N1c2} we display the results for this design, showing which structures are formed depending on the geometrical parameters $\cos\theta_{max}$ and $\gamma$. Comparing these results with Fig.~\ref{N1c1}, we observe that all three polyhedral shells can be assembled within the parameters explored. Differently from before, the structures form at all values of $\cos\theta_{max}$ considered, and the dominant structure is controlled primarily by the angle $\gamma$, with the stability of each structure approximately centered around its optimal angle. As observed before, structures with fewer particles are favored when the patch width increases: this is an entropic effect, as fully formed shells behave as an ideal gas of clusters, whose entropy increases with number density of clusters. Snapshots of the gas of icosahedral, snub cube, and snub dodecahedron are shown in the lower panels of Fig.~\ref{N1c2}.

Already from the results of Fig.~\ref{N1c2} we can assert that even the introduction of minimal patch colouring significantly improves the self-assembly process. In the next section we will look in detail the effects of different colouring patterns on the yield of self-assembly.

\subsection{Patch patterning}

\begin{figure*}
	\includegraphics{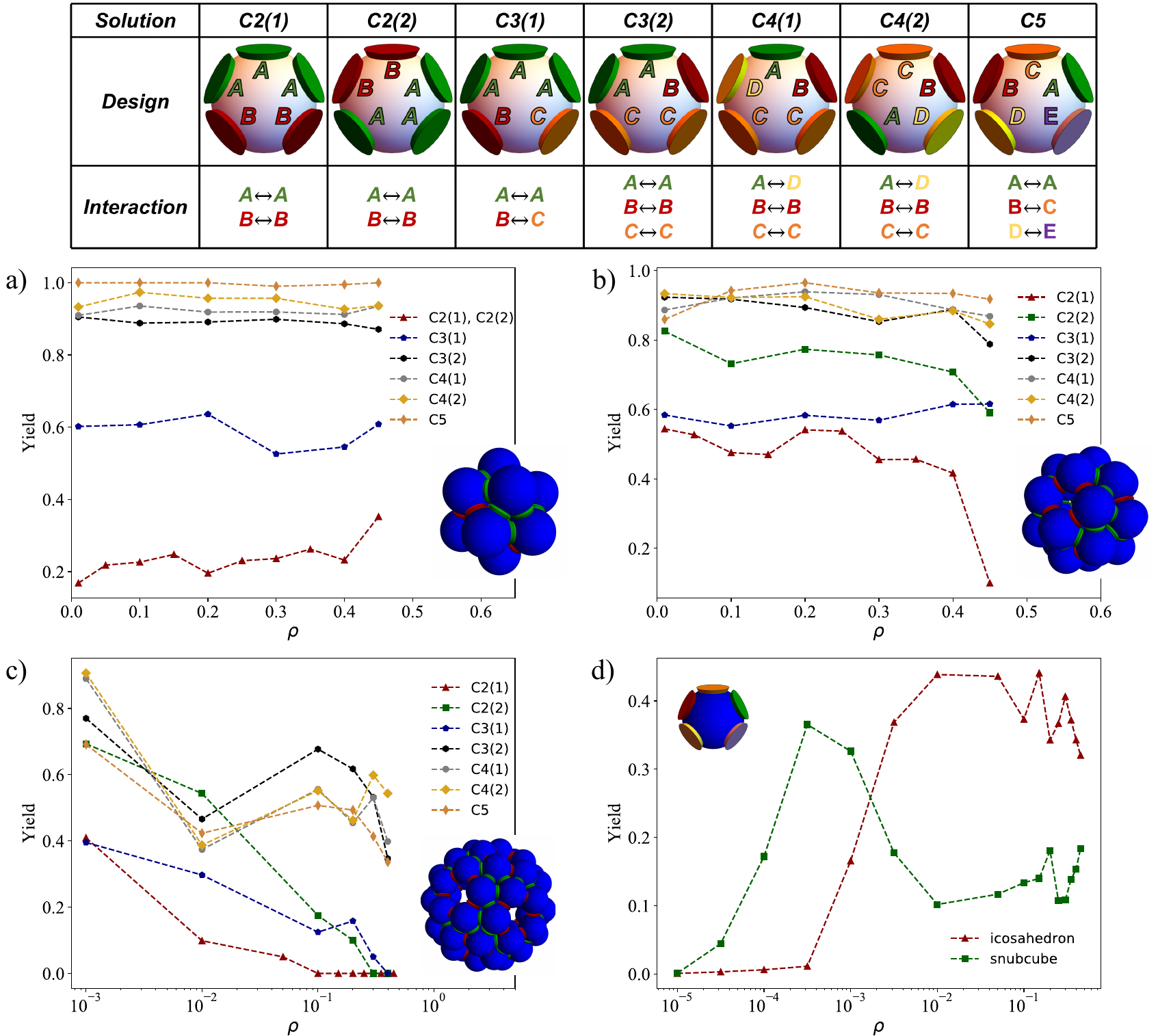}
	\caption{\label{Yield} Top: Graphic representation of the different one-component SAT solutions explored in this study, clarifying the respective patch coloring and interactions rules. Average yield of the icosahedron (panel a), snub cube (panel b) and snub dodecahedron (panel c) as a function of the density of patchy particles. These results were calculated with $\cos \theta _{max}=0.98$ and the in-plane angle, $\gamma $, was chosen to be the best for each structure. Thus, the icosahedron curve was calculated with $\gamma =60^\circ $, the snub cube with $\gamma =90^\circ $ and the snub dodecahedron with $\gamma =90^\circ $. Panel d) shows the average yield as a function of density for a design of five colours. Here, both curves were measured for the same system with $\gamma =85.45^\circ $ and $\cos \theta _{max}=0.947$. All results shown in this figure were calculated at $T=0.097$. Temperature ($T$) is expressed in units of $\varepsilon$ and $k_B=1$, while all lengths are in units of $\sigma$.
	}
\end{figure*}

The table in Fig.~\ref{Yield} contains seven different patch patterning designs, varying the number of colours and the interaction among them. As the number of colours increases, so does the complexity of the design. As such, SAT-assembly frameworks becomes essential to assign the colour interactions such that all desired target structures are possible. The name of the design expresses the number of colours used and, in parenthesis, different patterning choices: for example, $C4(1)$ and $C4(2)$ are two different designs with 4 colours. Each design is compatible with all three target structures, and we test them for three geometrical arrangements of the patches, differing for their angle $\gamma\in [60^\circ, 90^\circ, 108^\circ]$. It is important to note that for $\gamma=60^\circ$ all patches are geometrically indistinguishable, i.e. the angle between any two adjacent patches and the center of the particle is always $60^\circ$. For the other values of $\gamma$, instead, the only geometric symmetry of the particles is a vertical mirror plane as shown in Fig.~\ref{SAT}. This last symmetry can be broken by introducing patch colouring. By colour-symmetry we refer to the presence of symmetry elements (in our case the vertical mirror plane) which bring a specific design into self-coincidence after (eventually) exchanging the identity of the colours. For example, let's consider the design C3(1) in Fig.~\ref{Yield}: patches $(1,2,3)$ are green, patch $4$ is red, and patch $5$ is orange. The design has colour symmetry because after reflecting all patches through the vertical mirror plane, and after the following colour exchange $(\text{red}\leftrightarrow\text{orange})$, the design remains the same.

In the following we will investigate how colour-symmetry and the total number of distinct colours used affects the self-assembly yield.

In Fig.~\ref{Yield} we plot the density dependence of the yield for all designs and for $\gamma=60^\circ$ (panel a), $\gamma=90^\circ$ (panel b), and $\gamma=108^\circ$ (panel c).
We define the yield as the probability of finding a cluster corresponding to a specific structure. We count single particles as a cluster of size one, and any bonded particles as clusters of size two or above, depending on the number of particles bonded. For example, panel a) shows the probability of finding a cluster that forms an icosahedron, thus number of icosahedra formed over the total number of clusters. We stress the fact that we measure the equilibrium yield, i.e. the yield after long waiting times, as the implemented AVB biased moves allow the particles to rapidly form bonds regardless of the system density.

We can summarize the results of Fig.~\ref{Yield} with the following observations:\\
\noindent
\emph{1. Increasing the number of colours increases the yield.} Regardless of the target structure, and with few exceptions detailed below, the yield increases with the number of colours used. The increase is more significant for the first few colours added, while the yield's gain is more modest when the maximum number of colours is reached for a given number of species (in our case, the maximum number of colours is five times the number of different particle species).

This is due to the fact that the probability of creating an undesired bond, i.e. interaction between compatible patches which however creates a particle cluster whose topology (graph of all formed bonds) is not a subset of the target shape, decreases with increasing number of colours. For example, in the C2(1) design, the top patches can form three possible connections, and thus the probability that the top three green patches form a desired bond is one third, while for the two red bottom ones is one half. The probability of assembling the desired bonds to a central particle is then $(1/3)^3 (1/2)^2\sim 0.009$. As the number of colours increases so does the probability that a desired bond is formed. So for the C5 design, there is only one possible bonding partner for each patch, and so the only allowed bond topology is the one of the target structure.

\begin{figure*}[t]
	\includegraphics{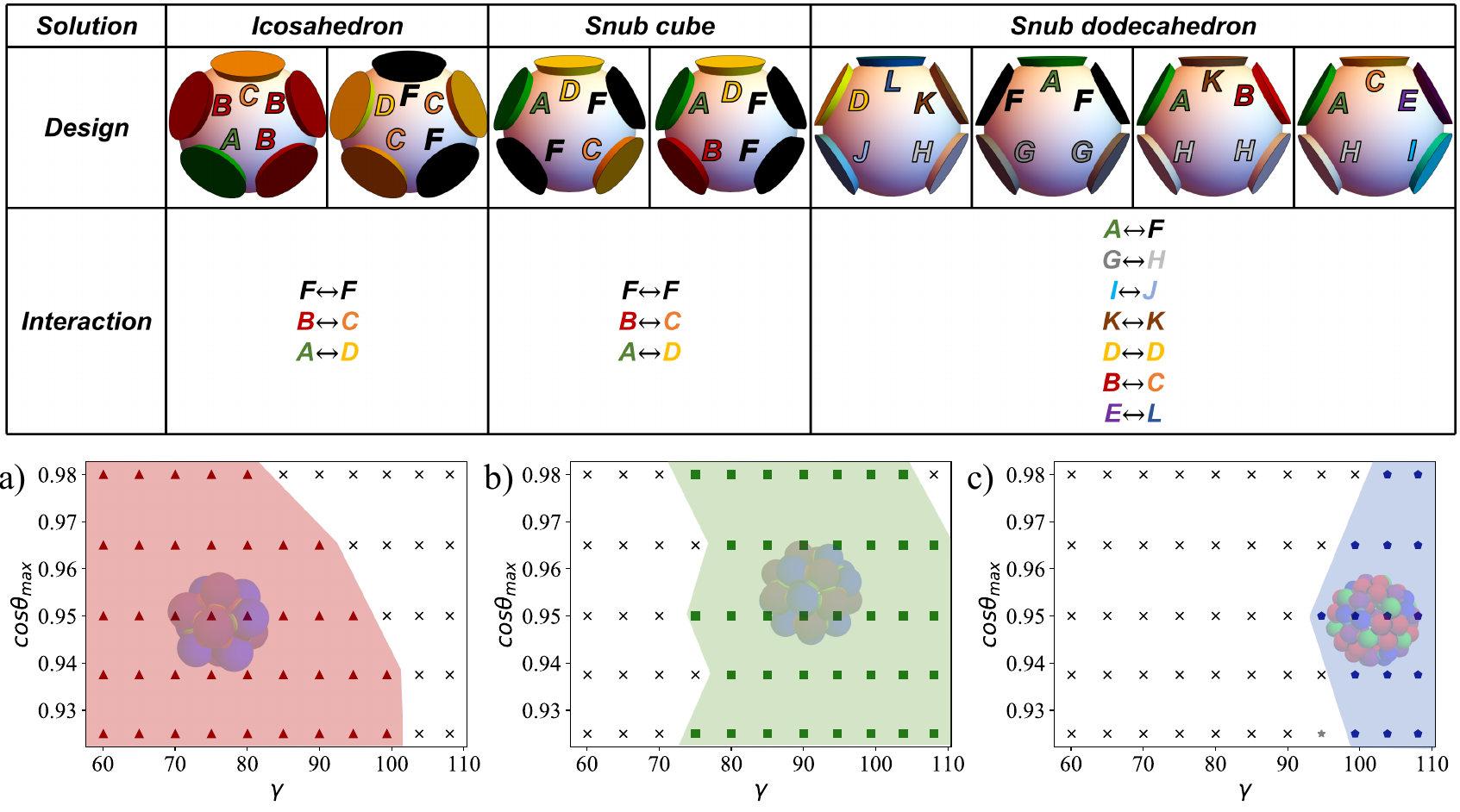}
	\caption{\label{Sol} Top Table: Graphic representation of the different multi-particle SAT solutions explored in this study, clarifying the respective patch colouring and interactions rules. Additional information regarding the designs are reported in the \textit {Supporting Information}. Panels (a), (b), and (c) show the most probable structure formed for different in-plane angles, $\gamma $, and for different patch width, $\cos \theta _{max}$, each for one specific SAT design. Panel (a) is based on a two species and five colours design, which allow the formation only of the icosahedron. Panel (b) is also based on a two species and five colours design, but which allow only the formation of the snub cube. Finally (c) describes a four species and twelve colours SAT solution for which a fully bonded configurations can be achieved only in the snub dodecahedron geometry. As for the previous figures, the coloured areas indicates regions of the parameter space where self-assembly of the desired structure is successful. Results were calculated with $T=0.097$ and $\rho =0.01$. Temperature ($T$) is expressed in units of $\varepsilon$ and $k_B=1$, while all lengths are in units of $\sigma$.
	}
\end{figure*}

\noindent
\emph{2. Decreasing the symmetry of the building block increases the yield}. While for $\gamma=60^\circ$ the particles have a five-fold symmetry axis, altering the $\gamma$ angle orients the particles, making all patches distinguishable. The best example are designs C2(1) and C2(2), which are indistinguishable for $\gamma=60^\circ$ (panel a), and whose yield increases significantly going from $\gamma=60^\circ$ (panel a) to $\gamma=90^\circ$ (panel b). Notice that the design C3(1) has a similar yield for $\gamma=60^\circ$ (panel a) and $\gamma=90^\circ$ (panel b). To understand this behaviour we next introduce the concept of \emph{colour symmetry}.

\noindent
\emph{3. Decreasing the colour symmetry of the patches increases the yield}. As mentioned before, all designs for $\gamma\neq 60^\circ$ have a single geometrical mirror plane, but patch colouring can maintain or break this symmetry. The designs that break the mirror-plane symmetry are \emph{chiral} designs: C2(2), C3(2), C4(1), C4(2), and C5. The designs that preserve the mirror-plane symmetry are \emph{achiral} designs: C2(1) and C3(1). Among the target structures, the icosahedron is an achiral assembly, while both the snub cube and the snub dodecahedron are chiral assemblies. We observe how in all cases chiral designs have higher yield compared to the achiral ones. This is true also for the icosahedron despite the lack of chirality in the target structure. For the snub cube and the snub dodecahedron we even observe that the yield of the chiral design C2(2) is higher than the achiral design C3(1) despite using less colours. Colouring reduces the symmetry of the target structure, which in turn reduces the number of degenerate structure that can form during the assembly. Controlling the colouring of the patches is thus an effective strategy to increase the yield of the assembly.

\noindent
\emph{4. The yield has only a weak density dependence.} We observe that the yield of the different structures is constant with density. This means that clusters are fully formed in our thermodynamic conditions, and have ideal-gas behaviour. Only at high densities the yield starts decreasing, in correspondence with inter-cluster interactions and possibly a phase change to a liquid. Interestingly, we also observe that large structures like the snub dodecahedron, which is composed of $60$ particles, is more easily assembled in very dilute systems where larger shells have more space to grow without influencing each other.

\noindent
\emph{5. Frustrated designs allow to target different structures depending on thermodynamic conditions.} In Fig.~\ref{Yield}\textit{d}  we plot the yield for the frustrated C5 design in which $\gamma=85.26^\circ$, thus it does not ideally satisfy any of the structures discussed previously but --- given the used patch width --- the icosahedron and snub cube can still be assembled.
The in-plane angle is close to the ideal angle for the assembly of the snub cube, which is indeed the most stable structure at low densities. But increasing the density we observe a switch to the smaller icosahedral shells. While the icosahedron has a much larger free energy of formation compared to the snub cube (due to the angle being unfavourable to the icosahedron), at high density this is compensated by the translational free energy (i.e. the ideal gas free energy), which is higher for the icosahedron (having a higher number density of clusters compared to the snub cube). The possibility to switch between different target structures as a function of external control parameters is a property of experimental interest.

We note on passing that several thermodynamic properties, including the critical micelle concentration, have been evaluated for all these single-species designs. The results are discussed in detail in
the \textit{Supporting Information}.

\subsection{SAT-assembly selection: Eliminating competing structures}

In the previous examples, control over the target structure was obtained either geometrically (by changing the angles $\gamma$ and $\theta_\text{max}$ as in Fig.~\ref{N1c2}), or thermodynamically (by changing the density as in Fig.~\ref{Yield}d). Here we demonstrate that SAT-assembly allows to encode structure selection directly into the patch colouring. In particular we look for designs that satisfy only the icosahedron but not the snub cube and snub dodecahedron, and vice-versa. We find that the mutual exclusion of all three target structures requires at least two particle species for the snub cube and icosahedron, but four species for the snub dodecahedron. Here we show results with two species ($N_p=2$) and five colours ($N_c=5$) for the icosahedron and snub cube and four species ($N_p=4)$ and twelve colours ($N_c=12$) for the snub dodecahedron, even if other solutions with different number of colours also exist. The results with our selected designs are shown in Fig.~\ref{Sol}. For the two species design we highlight that a mutually exclusive selection of the target structures can be achieved via non trivial patch colour ordering using SAT.

Comparison with the 1-specie solution (see Fig.~\ref{N1c2}) shows that suppression of the competing structure significantly enlarges the range of parameters where a certain shell is formed. Finally, the absence of competing structures also increases the yield to almost $100\%$ in some regions of the parameter space for the icosahedron and to $50\%$ for the snub cube and snub dodecahedron (see \textit{Supporting Information}).

Incidentally, we note that targeting only the snub dodecahedron is complicated by the fact that all two species designs found by SAT are compatible with the formation of icosahedron or snub cube structures from only one of the two species, practically pre-emptying the formation of the target structure. Using the SAT framework, we proved that no solution exist with only 2 particle species. As such, the number of species needs to increase in order to find a suitable design that satisfies the snub dodecahedron while completely excluding the others. We found that a four species design satisfies the constraint. Indeed, using SAT, one can calculate a solution for the snub dodecahedron and then check if it (or any of its subsets) also satisfies the other structures. If so, this solution is excluded and a new one is generated until all solutions are exhausted. Therefore, it is possible that SAT designs with fewer species satisfy all constraints, but require significant computational resources to find them using this method.

We note that the range of parameters where the snub dodecahedron is formed enlarges, but not as much as in the case of the icosahedron or of the snub cube. In fact, the snub dodecahedron is not as robust to geometrical frustrations as the other structures. As $\gamma$ approaches $90^\circ$, snub cubes start forming due to geometrical incompatibilities, but given the colouring they never fully close. These almost complete structures require a large amount of time to break.

\subsection{Short time kinetics}\label{sec:kinetics} 

Here we explore the effect of colouring on the short time kinetics of the self-assembly process using Molecular Dynamics. We use a continuum version of the Kern-Frenkel potential and focus on the assembly process of the icosahedron. We restrict all results below to $\rho=0.1$ and $T=0.097$, as used in the previous sections. We also fix $\cos \theta_{max}=0.97$ and $\gamma=60^\circ$ to more easily target the icosahedral shell. More detailed information regarding the Molecular Dynamics simulations is presented in the \textit{Material and Methods} section.

We focus on 10 different SAT solutions for the icosahedral shell and test for each of them the associated short time kinetics. The first five correspond to solutions that have one specie and an increasing number of colours: according to the nomenclature of Fig.~\ref{Yield}, they are C1, C2(2), C3(2), C4(2), and C5.
	The remaining 5 solutions increase the number of species and use the corresponding maximum number of colours, in order C10 (2 species), C15 (3 species), C20 (4 species), C30 (6 species), and C60 (12 species).

\begin{figure}[t]
	\begin{center}
		\includegraphics{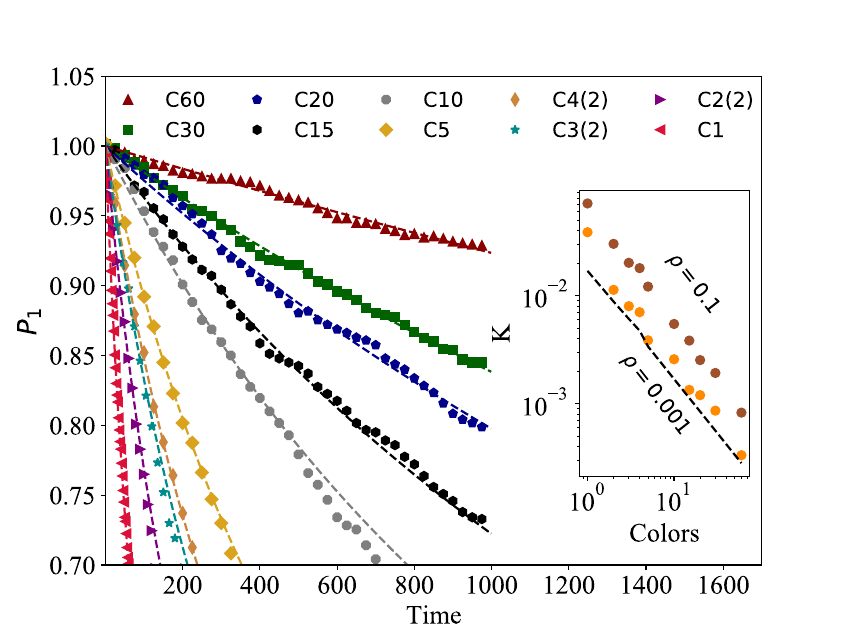}
	\end{center}
	\caption{\label{kinetics} Short time kinetics of the icosahedron assembly for different numbers of species and colours. In the main plot is shown the time evolution of the fraction of monomers, $P_1(t)$, for short times using different designs with multiple species and colours. The lines are fits using $P_1(t)=1/(1+K\rho t)$, with $K$ as a fitting parameter and $\rho=0.1$, for the short time scale of the kinetics (the fit is constrained to the range where $P_1$ is higher than $80\%$). In the inset is shown the aggregation rate as a function of the number of colours for two different densities, $\rho=0.1$ and $\rho=0.001$. We also show a theoretical line given by $K=8\pi D(\sigma+\delta)\chi^2\mathcal{C}$, with $D=0.1$. All results were calculated using $T=0.097$ and averaged over 5 independent samples.
	}
\end{figure}

Figure~\ref{kinetics} plots the fraction of monomers (non-bonded particles), $P_1$, as a function of time for the 10 different explored SAT solutions. We observe that adding more colours and species slows down the short time kinetics of the system. Thus, it will take longer to reach the equilibrium state. To quantify this slow down we use the Smoluchowski coagulation equation to calculate the short time aggregation rates~\cite{Sciortino2009a, Ghofraniha2009}, and extend it to take into account the effect of colouring. In particular, we write:

\begin{equation}
\frac{dP_1}{dt}=-K \rho P_1(t)P_1(t) \ \ .
\label{eq:pdit}
\end{equation}

We compute the aggregation rate K from the short term fits of the solution of Eq.~\ref{eq:pdit}, 

\begin{equation}
P_1(t)=\frac{1}{1+K \rho t} \ \ ,
\label{eq:pditsol}
\end{equation}

\noindent shown as dashed curves in Fig.~\ref{kinetics}. The inset of Figure~\ref{kinetics} shows the aggregation rate $K$ plotted as a function of the number of colours. For isotropic particles, the Smoluchowski coagulation theory predicts $K=8\pi D\sigma$, where $D$ is the diffusion coefficient, and $\sigma$ is the diameter of the particle. To extend this to coloured patchy particles we write:

\begin{equation}
K=8\pi D\sigma\chi^2 \mathcal{C} \ \ ,
\end{equation}

\noindent where $\chi=(1-\cos{\theta_{max})/2}$ is the fraction of the particle surface covered by a single patch, and $\mathcal{C}$ is the number of bond combinations among two particles allowed by the interaction matrix.
	For the C1 design (all patches having the same colour) there are $\mathcal{C}_\text{C1}=25$ possible bond combinations between two particles (each patch in the first particle can interact with each patch on the second particle). For C2(2), as seen in Fig.~\ref{Yield}, 3 green patches on the first particle can interact with 3 green patches on the second particle (for a total of 9 combinations), while the 2 red patches on the first particle interact only with 2 red patches on the second particle (for a total of 4 combinations), giving $\mathcal{C}_\text{C2(2)}=13$. With similar considerations one finds that $\mathcal{C}_\text{C3(2)}=9$, $\mathcal{C}_\text{C4(2)}=7$, and $\mathcal{C}_\text{C5}=5$. Full coloured solutions ($N_c=5 N_p$) follow the simple rule
	$\mathcal{C}_{N_p}=5/N_p=25/N_c$, i.e. the aggregation rate scales as the inverse number of colours used.

	In the inset of Fig.~\ref{kinetics} we plot the aggregation rate $K$ measured for two different densities ($\rho=0.1$ and $\rho=0.001$) and the theoretical line $K=8\pi D(\sigma+\delta)\chi^2\mathcal{C}$. We observe that the simulations approach the theoretical prediction as the system becomes more diluted. We also see that even at high densities, the aggregation rate obeys the scaling law $K\approx N_c^{-1}$.

\section{Conclusions}

In this work we have explored design principles that can guide the formation of complex target structures, and applied them to regular polyhedral shells of valence five, i.e. the icosahedron, the snub cube, and the snub dodecahedron. In particular we have explored how the inter-particle interactions can encode a predetermined target structure, a strategy that has many counterparts in the biological world, as for example in the self-assembly of virus capsids.

Choosing the correct inter-particle designs (i.e. patch colouring) is a complex optimization problem that we solve by encoding the bond topology in a set of satisfiability equations. This approach, named SAT-assembly, not only efficiently searches the space of possible designs for solutions which have the target structure as an energy minima, but also allows to explicitly enforce the non-satisfiability of competing structures as a way of avoiding the formation of kinetic aggregates.

Starting from solutions which target at the same time all the structure of interest, we have explored the effects of patch geometry, patch colouring, and patch patterning on the aggregate's yield. We find that the symmetry of the building blocks plays a key role in determining the yield of the final structure, with chiral designs consistently producing high yields for all structures considered. Introducing frustration by altering the patch geometry from the ideal one, is a promising strategy to produce designs with target structures that depend on external conditions.

We used SAT to selectively target one structure while excluding competing ones. For this, we increased the complexity of the design to two particle species and five patch colours for the icosahedron and snub cube, and four species and twelve colours for the snub dodecahedron. For these designs we observe that the yield significantly increases (to almost $100\%$ in multiple regions of the parameter space) and we also find a wider parameter range where it is possible to successfully assemble these structures. Thus, using SAT to suppress competing structures is a quite promising strategy for high-yield assembles.

We have also explored the short time kinetics of the icosahedral shell for different designs that vary in number of colours and species. Using the Smoluchowski coagulation equation we show that the short time kinetics slows down with increasing number of colours, with an inverse proportionality law. Since the static yield saturates quickly with the number of colours, an optimal number of colours can be found that guarantees high yields in accessible experimental times. In this context, our assembly rules acquire even more significance, because they show how to optimize the yield with changing colour arrangement but without changing the overall number of colours (e.g. preferring chiral arrangements over non-chiral ones). For the respective capsid designs considered here, we observed nearly perfect yield with already just one specie, making the use of more of them redundant. We notice also that increasing the number of different colours lowers the assembly temperature, potentially favouring --- due to irreversible bonding --- the formation of misfolded clusters. Hence it is quite important to select in experiments the  model that gives the required yield with the lowest number of colours. However, for other systems with fewer symmetries than the capsids considered here, the trade-off between number of species, desired yield, and assembly kinetics might result in a larger number of species to be preferred \cite{bohlin2022}.

Lastly, we also explored the phase diagram of these patchy particles designed with SAT. We find a non-monotonous behavior of the average potential energy as a function of density, as is typical of self-assembly systems \citep{Sciortino2009}, where an ideal gas phase is found at very low densities, followed by a gas phase of clusters near the energy minima, ending at the liquid phase at high densities.

All the state diagrams shown in the main text were calculated using the same densities and temperatures. We expect that as long as the system remains in the same thermodynamic phase, the yield will remain almost constant. The results shown in Fig.~\ref{N1c2} support this expectation, since there the density was the control parameter. Of course, repeating the calculation with significantly higher temperatures will prevent the observation of clusters, since the system will be well bellow the critical miscelle concentration. Similarly, significantly lower temperatures can lead to large metastable clusters or even percolating ones that reduce the yield of the small shells. Thus, for the finite size shells, the optimal yield should lie within an intermediate temperature range corresponding to the ideal gas phase of fully formed aggregates.

One of the possible pathways of realizing these designs experimentally is through 3D DNA nanomaterials, in particular wireframe DNA origami. Previous studies have successfully shown the versatility of these building blocks in assembling a wide array of structures \citep{Mosayebi2017, Lee2022, Jun2021, Rothemund2006}. Recent work has achieved capsid assembly from 3D DNA origamis using shape complemntary building blocks, akin to fitting puzzle pieces which bind at fixed prescribed angle \citep{sigl2021programmable}.  We note that our designs of coloured patches can also be realized through use of complementary strands, where compatible patch colours correspond to complementary DNA strands that functionalize the wireframe nanostructure to act as a patchy particle with selective spatial bonding and tune the interactions accordingly \citep{Biancaniello2005, Wang2015}. Such wireframe designs are expected to be easier to design than shape complementary origami approach, and furthermore can be used to a reconfigurable system, which can form different target shapes based on which patch colours (DNA strands) are available. We also argue that our results support such approach not only due to the high yields observed in simulations, but also due to the robustness of the structures formed to flexibility of the bonds, which is characteristic of DNA bonds \citep{Meulen2015, Meulen2015, Geerts2010}. Although we used an idealized patchy particle model,  ongoing experimental DNA-origami results indicate that interaction designs based on simulations with this model can predict the structures obtained from the assembly of polyhedral DNA wireframes in experiments. Aside from flexible DNA wireframe origami, this type of design can be easily extended to proteins \cite{Jie2021}, or colloidal particles with brushes \cite{McMullen2022}

For simplicity, we allowed for complementary bounding between patches. If such is not possible, adding more particle species is a simple way to introduce the non-complementary bonding constraint. From what was shown here, we do not expect that such change should impact significantly the results, especially since it usually reduces the symmetry of the building blocks which promotes higher yields.

While we focused here in particular on three different shell designs, our design method could be straightforwardly applied to other capsid geometries, thus providing a computational pipeline of self-assembled shell designs for their possible nanotechnology applications.

\section{Methods}

We consider a system composed of $N$ patchy particles in a cubic box of length $L$. Particles are characterized by a hard core of diameter $\sigma$ with five patches on its surface. The patches interact through the Kern-Frenkel potential \citep{Kern2003, bol1982monte}:

\begin{equation}
Vpp(\boldsymbol{r}_{ij}, \boldsymbol{\hat{r}}_{\alpha, i}, \boldsymbol{\hat{r}}_{\beta, j})=V_{SW}(r_{ij})f(\boldsymbol{r}_{ij}, \boldsymbol{\hat{r}}_{\alpha, i}, \boldsymbol{\hat{r}}_{\beta, j})
\end{equation}

where $i$ corresponds to a given particle and $\boldsymbol{r}_{i}$ its center of mass. Thus, $\boldsymbol{r}_{ij}$ is the distance between particles $i$ and $j$. $\boldsymbol{r}_{\alpha, i}$ denotes the position of patch $\alpha$ of particle $i$. $V_{SW}$ is an isotropic square-well of range $\sigma + \delta_{\alpha,\beta}$ and depth $\varepsilon_{\alpha,\beta}$, the hat symbol indicate unit vectors and $f$ is the orientation-dependent modulation term that takes the form:

\begin{equation}
\label{KF}
f(\boldsymbol{r}_{ij}, \boldsymbol{\hat{r}}_{\alpha, i}, \boldsymbol{\hat{r}}_{\beta, j})=
\begin{cases}
1 & \text{if $\begin{aligned}
	\text{$\boldsymbol{\hat{r}}_{ij} \cdot \boldsymbol{\hat{r}}_{\alpha, i} > \cos     \theta^{max}_{\alpha \beta}$} \\
	\text{$\boldsymbol{\hat{r}}_{ji} \cdot \boldsymbol{\hat{r}}_{\beta, j} > \cos     \theta^{max}_{\alpha \beta}$}
	\end{aligned}$ } \\
0 & \text{otherwise}
\end{cases}
\end{equation}

\begin{figure}[t]
	\begin{center}
		\includegraphics[width=0.25\textwidth]{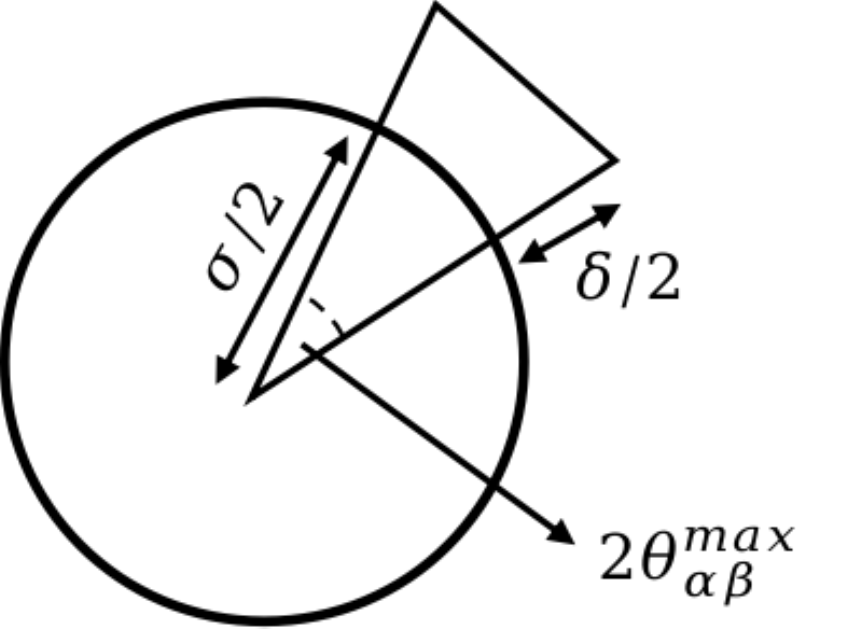}
	\end{center}
	\caption{\label{cosmax} Schematic representation of the Kern-Frenkel portential using the cross section of a particle.
	}
\end{figure}

With this formulation patches are represented by a cone starting from the center of mass of the particle and reaching $\sigma + \delta_{\alpha,\beta}$, while the width is controlled by $\theta^{max}_{\alpha \beta}$ (as shown in Fig.~\ref{cosmax}). This potential has been extensively used to study systems of patchy particles \citep{Rovigatti2018}. For simplicity, we consider the parameter range where it is only possible to form one bond per patch. In the following, $\sigma$ provides the unit of length and $\varepsilon_{\alpha, \beta}$ the unit of energy. Temperature ($T$) is also expressed in units of $\varepsilon_{\alpha, \beta}$ and $k_B=1$.

For the following results we considered Monte Carlo (MC) simulations with two possible moves, roto-translations and aggregation-volume-bias~\citep{Rovigatti2018}. The first attempts a simple rotation and translation of a random particle along a random (radial or angular) direction. The second, attempts to move a random particle into the vicinity of another such that a bond is formed between the two. To not break ergodicity, the inverse move can also be performed where a random bond between two particles is broken. We performed simulations in the $NVT$ assemble to explore the assembly of our desired shells. All results shown bellow were averages over simulations with or more than $10^8$ MC time-steps. For all, we considered $N=480$ and $\delta_{\alpha, \beta}=0.2$. The simulations start with particles randomly generated in the box with random orientations. Unless stated otherwise, all results are averages over $10$ independent samples.

We follow the same SAT formulation as the one in Ref.~\citep{Russo2022}. We consider that each patch can have a given colour between $1\leq x_c\leq N_c$, where $N_c$ is the total number of colours. These colours can be distributed onto the patches in specific arrangements, each unique sequence can be considered a particle specie, thus $1\leq x_p\leq N_p$, where $N_p$ is the total number of species. SAT is then used to find if a given combination of $N_p$ and $N_c$ can satisfy a given polyhedral shell, e.g. if it satisfies all the topological constraints, and a solution/design is calculated which can be used to prepare the composition of the system. In the \emph{Supporting Information} we go into more detail on the different constraints (clauses) used in SAT.

In Fig~\ref{SAT} we show schematically one of the patchy particles of valence five used and the different shells assembled. There are three possible polyhedron shells that can form and fully close, depending on the parameters and SAT solution used: the regular icosahedron, the snub cube and the snub dodecahedron. The positions of the five patches, for the case of the icosahedron, in the orthonormal base associated with the patchy particle, are given as:

\begin{equation}
\label{patch}
\begin{aligned}
\textbf{p}_1=(-\sqrt{\frac{\varphi+2}{5}}, 0, \frac{\varphi-1}{\sqrt{3-\varphi}}) \\
\textbf{p}_2=(\frac{1-\varphi}{2}\sqrt{\frac{\varphi+2}{5}}, -\frac{\varphi}{2}, \frac{\varphi-1}{\sqrt{3-\varphi}}) \\
\textbf{p}_3=(\frac{\varphi}{2}\sqrt{\frac{\varphi+2}{5}}, -\frac{1}{2}, \frac{\varphi-1}{\sqrt{3-\varphi}}) \\
\textbf{p}_4=(\frac{\varphi}{2}\sqrt{\frac{\varphi+2}{5}}, \frac{1}{2}, \frac{\varphi-1}{\sqrt{3-\varphi}}) \\ 
\textbf{p}_5=(\frac{1-\varphi}{2}\sqrt{\frac{\varphi+2}{5}}, \frac{\varphi}{2}, \frac{\varphi-1}{\sqrt{3-\varphi}}) .
\end{aligned}
\end{equation}

\noindent where $\varphi$ is the golden ratio. To form the other structures one can increase the in plane angle, $\gamma$, between $\textbf{p}_4$ and $\textbf{p}_5$. We do that by using $\textbf{p}_3$ as an axis of rotation for $\textbf{p}_4$ and $\textbf{p}_1$ as an axis of rotation for $\textbf{p}_5$. We multiply $\textbf{p}_4$ and $\textbf{p}_5$ by the respective rotation matrix, Eq.~\ref{rotmat}, where $\textbf{p}_{x,\alpha}$ refers to the rotation axis vector and $\alpha$ the vector index. The angle of rotation $\theta$ is used to increase the in plane angle $\gamma$. At $\theta=0$, the in plane angle is $\gamma=60^\circ$, while at $\theta\approx-46.5$ (and $\theta\approx46.5$ for $\textbf{p}_{5}$) the in plane angle reaches the maximum value used of $\gamma\approx108^\circ$.

\begin{figure*}
	\begin{align}
	\label{rotmat}
	\begin{bmatrix}
	\sqrt{\textbf{p}_{x,1}}+(1-\sqrt{\textbf{p}_{x,1}})\cos\theta & \textbf{p}_{x,1}\textbf{p}_{x,2}(1-\cos\theta)-\textbf{p}_{x,3}\sin\theta & \textbf{p}_{x,1}\textbf{p}_{x,3}(1-\cos\theta)+\textbf{p}_{x,2}\sin\theta \\
	\textbf{p}_{x,1}\textbf{p}_{x,2}(1-\cos\theta)+\textbf{p}_{x,3}\sin\theta & \sqrt{\textbf{p}_{x,2}}+(1-\sqrt{\textbf{p}_{x,2}})\cos\theta & \textbf{p}_{x,2}\textbf{p}_{x,3}(1-\cos\theta)-\textbf{p}_{x,1}\sin\theta \\
	\textbf{p}_{x,1}\textbf{p}_{x,3}(1-\cos\theta)-\textbf{p}_{x,2}\sin\theta & \textbf{p}_{x,2}\textbf{p}_{x,3}(1-\cos\theta)+\textbf{p}_{x,1}\sin\theta & \sqrt{\textbf{p}_{x,3}}+(1-\sqrt{\textbf{p}_{x,3}})\cos\theta
	\end{bmatrix}
	\end{align}
\end{figure*}

Using SAT we can find a minimal design that satisfies all three structures. For example, it is possible to consider the case that is shown in Fig.~\ref{SAT}, where we only use one specie (blue) of particles and two colours (green and red) for patches. In this design, green patches only interact with green and red with red. If the particles follow this colouring and interactions then all three structures can in principle form. The SAT solution that leads to this design is not necessarily the only one that satisfies all structures but SAT only provides one solution at a time for the constraints provided. Nonetheless, SAT is flexible enough, such that, we can provide this solution found as a new constraint and thus avoid a previous solution altogether. This leads to new solutions (translating into different particle designs). This process can be iterated until all solutions are exhausted.

\begin{figure}[t]
	\begin{center}
		\includegraphics[width=0.475\textwidth]{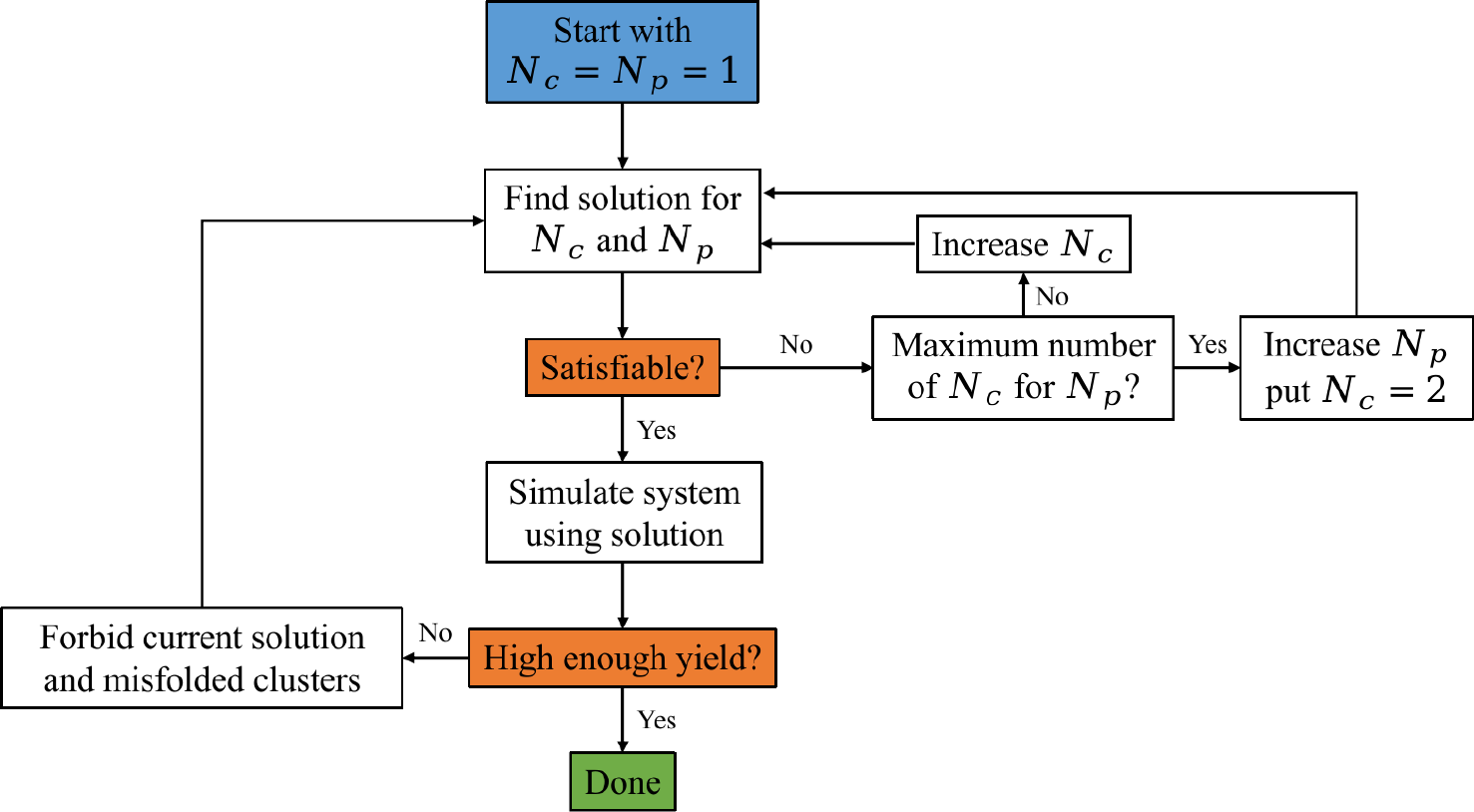}
	\end{center}
	\caption{\label{diagram} Diagram of the SAT pipeline.
	}
\end{figure}

One of the advantages of the SAT algorithm is that due to its high efficiency it can be easily integrated into a simulation pipeline to quickly develop a design that excludes the maximum number of shells. Figure \ref{diagram} shows the relevant steps of this pipeline. One starts by using SAT to calculate a solution that satisfies the targeted shell. Then the system is simulated using this design to find new misfolded shells different from the target one. These misfolded shells are then added to SAT and excluded from the new solution. Thus, those misfolded shells will no longer form. This process can be iterated until the yield of misfolded shells is negligible. For the geometry of patchy particles presented in this article, the three main closed shells are the ones in Fig.~\ref{SAT}. For the range of parameters explored, the misfolded shells have very low yields or are constrained to limits of the parameter space (wide patch widths). Thus, in the present case, the path in Fig.~\ref{diagram} coincides the straight line the blue to the green box.

The code implementing the SAT-assembly pipeline for polyhedral shells is available at https://github.com/deppinto/PatchyParticles.

The short time assembly simulations were performed using a Molecular Dynamics method with a generalization of the Kern-Frenkel potential between patches~\cite{Rovigatti2022}. Patchy particles feel a mutual repulsion modelled through a WCA interaction:

\begin{equation}
U_{ij}(r) =\left\{
	\begin{array}{l l}
	4\varepsilon \left[ \left( \frac{ \sigma }{r} \right)^{12} - \left( \frac{ \sigma }{r} \right)^{6} + \frac{1}{4}\right] & r \le 2^{\frac{1}{6}} \sigma\\
	0 & r > 2^{\frac{1}{6}} \sigma
	\end{array} \right.
\label{eq:wca}
\end{equation}

\noindent where $r$ is the distance between particle centers, $\varepsilon$ is the energy scale and $\sigma$ is the particle diameter. Thus, energy is in units of $\varepsilon$ while lengths are in units of $\sigma$.

The patch-patch interaction is a square-well-like attractive potential modulated by an orientation-dependent function. The range of the interaction is given by $\delta$, while its angular width by $\cos \theta_{\rm max}$. The interaction between patch $i$ on particle $\alpha$, identified by the unit vector $\hat{\alpha}_i$, and patch $j$ on particle $\beta$, identified by $\hat{\beta}_j$, is given by:

\begin{multline}
V_{\rm pp}(\vec{r_{\rm pp}}, \hat{\alpha}_i, \hat{\beta}_j) = \\  -\varepsilon \exp\left(-\frac{1}{2} \left( \frac{r_{\rm pp} - \sigma_c}{\delta} \right)^{10}\right) \Omega(-\hat{r}, \hat{\alpha_i}) \Omega(\hat{r}_{\rm pp}, \hat{\beta_j})
\end{multline}

\noindent where $\vec{r}_{\rm pp} = \vec{r}_\alpha - \vec{r}_\beta$, $r_{\rm pp} = |\vec{r}_{\rm pp}|$, $\hat{r}_{\rm pp} = \vec{r}_{\rm pp}/r_{\rm pp}$ and $\Omega$ is a steep modulating function that takes into account the orientation of a patch with respect to the unit vector connecting the center of the particles and takes the following form:

\begin{equation}
\Omega(\hat{r}, \hat{\gamma_k}) = \exp\left(-\frac{1}{2} \left( \frac{1 - \hat{r}\cdot\hat{\gamma}_k}{1 - \cos \theta_{\rm max}} \right)^{20}\right).
\end{equation}

We set $\delta = 0.2$ and $\cos \theta_{\rm max} = 0.97$ so that only one bond can form per patch. We used the oxDNA package~\cite{Rovigatti2015, poppleton2023oxdna} to simulate the patchy particle system described above. We only focused on the icosahedral shell, so the patches were located at the positions given by Eq.~\ref{patch}.

\section{Acknowledgements}

The authors acknowledge all the financial support from the European Research Council Grant DLV-759187. This result is part of a project that has received funding from the European Research Council (ERC) under the European Union’s Horizon 2020 research and innovation programme (Grant agreement No. 101040035) (to P\v{S}).

\renewcommand\thefigure{S\arabic{figure}}
\setcounter{figure}{0}

%

\section{Supporting Information}

\subsection{Structure details}

To map the patchy particle design into a SAT problem it is necessary to translate it into boolean variables and then impose constraints such that the structures in Fig.~1 are formed.

The boolean variables can be divided into four major groups. The first group is the colour interaction variables, $x_{c_i,c_j}^{int}$, where $c_i$ and $c_j$ are the colour of particle $i$ and $j$ respectively. If this variable is true then colours $c_i$ and $c_j$ interact and can form a bond, otherwise not. There are a total of $(N_c)(N_c+1)/2$ of these variables. The second group is the patch colouring variable, $x_{p,s,c}^{pcol}$, where $p\in [1, N_p]$ refers to particle species, $s\in [1, V]$ to patch number and colour $c\in [1, N_c]$. If true, particle specie $p$ has the patch number $s$ of colour $c$. There are $N_pVN_c$ of these variables. Then the placement variables, $x_{l,p,o}^{L}$, where $l\in [1, L]$ refers the position of a particle in the polyhedron, $p\in [1, N_p]$ to particle specie and orientation $o\in [1, R]$. If true, a particle of species $p$ occupies position $l$ in the polyhedron according to orientation $o$. There are $N_pLR$ of these variables. Lastly, there is an auxiliary variable, $x_{l,s,c}^{A}$. If true, the particle in position $l$ is oriented such that the patch $s$ has a colour $c$. There are $VLN_c$ such variables.

The orientation mapping is given in Table I, while Tables II, III and IV present the polyhedron topology map for the icosahedron, snub cube and snub dodecahedron respectively. The patches are labeled as in Fig.~1. For the icosahedron, all patches are indistinguishable and thus they can be mapped with Table I.

There are seven main groups of clauses solved by SAT. The first guarantees that each colour can only interact with only one other colour:

\begin{equation}
C^{int}_{c_i,c_j,c_k}=\neg x_{c_i,c_j}^{int} \vee \neg x_{c_i,c_k}^{int}
\end{equation}

The second ensures that patch number $s$ of particle specie $p$ will have exactly one colour only:

\begin{equation}
C^{pcol}_{p,s,c_k,c_l}=\neg x_{p,s,c_k}^{pcol} \vee \neg x_{p,s,c_l}^{pcol}
\end{equation}

The third guarantees that position $l$ is occupied by exactly one particle specie with one orientation:

\begin{equation}
C^{L}_{l,p_i,o_i,p_j,o_j}=\neg x_{l,p_i,o_i}^{L} \vee \neg x_{l,p_j,o_j}^{L}
\end{equation}

The fourth enforces that the neighboring positions $l_i$ and $l_j$ connected by the patches $s_i$ and $s_j$ (given by the tables bellow) have colours in those patches, $c_i$ and $c_j$, which interact:

\begin{equation}
C^{lint}_{l_i,s_i,l_j,s_j,c_i,c_j}=\neg x_{l_i,s_i,c_i}^{A} \vee \neg x_{l_j,s_j,c_j}^{A} \vee x_{c_i,c_j}^{int}
\end{equation}

The fifth ensures that for a position $l$ that is occupied by particle specie $p$ with orientation $o$, the patch $s$ has the right colour attributed to it:

\begin{equation}
\begin{split}
C^{LS}_{l,p,o,c,s}= & ( \neg x_{l,p,o}^{L} \vee \neg x_{l,s,c}^{A} \vee x_{p,\phi_o(s), c}^{pcol} ) \\ 
& \wedge ( \neg x_{l,p,o}^{L} \vee x_{l,s,c}^{A} \vee \neg x_{p,\phi_o(s), c}^{pcol} )
\end{split}
\end{equation}

The two last groups define multiple clauses each, the first enforces that all particle species are used, while the second enforces that all colours are used:

\begin{equation}
\forall p\in [1, N_p]:C_p^{all p.}= \underset{\forall l \in [1,L], o\in [1,R]}{\bigvee} x_{l,p,o}^L
\end{equation}

\begin{equation}
\forall c\in [1, N_c]:C_c^{all c.}= \underset{\forall p \in [1,N_p], s\in [1,V]}{\bigvee} x_{p,s,c}^{pcol}
\end{equation}

\subsection{Thermodynamic properties}

\begin{figure*}[!t]
	\includegraphics{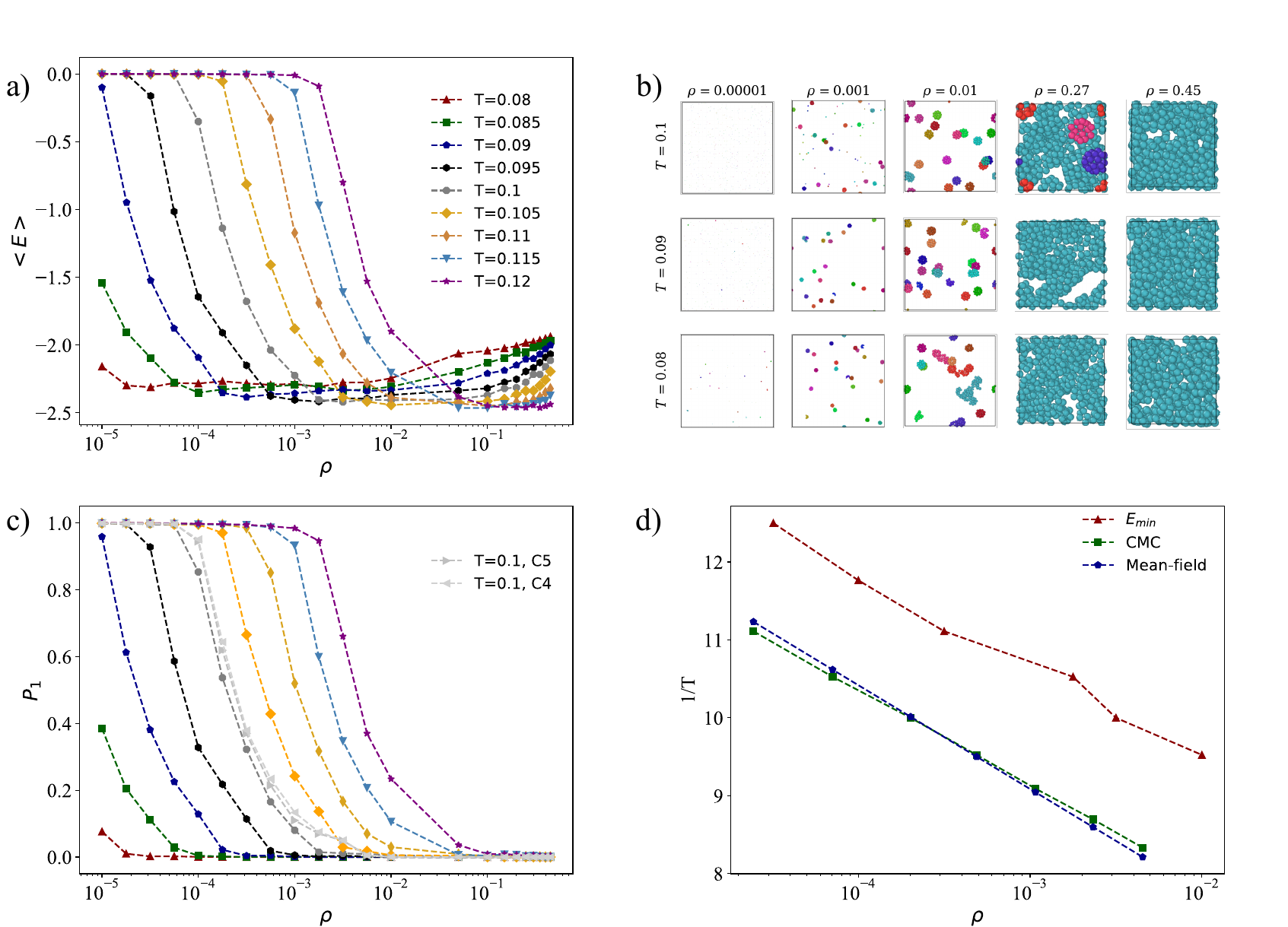}
	\caption{\label{Energy} a) Average potential energy per particle as a function of density for different isotherms. For clarity, the x-axis is in log-scale. We observe a non-monotonic behaviour of the average energy characteristic of self-assembly systems. b) Frontal snapshots of the system for different densities and temperatures. Images were obtained with OVITO, where the colours represent particles that belong to the same cluster. For low densities, some colours repeat themselves even though particles are not bounded due to the large number of unbounded particles which count as clusters of size one. c) Fraction of monomers as a function of the density. The color coding is the same as in a). Two new curves were added at $T=0.1$, with a different solution (as shown in Fig.~4). d) Inverse temperature as a function of density for the point at which the fraction of monomers is equal to $50\%$ (critical micelle concentration). A theoretical curve is drawn to estimate the CMC, it is calculated using Eq.~\eqref{meanfield}. $E_{min}$ corresponds to the energy minimum in a). All results were simulated with the one specie and two colour design (except for the two curves in the top right plot), with $\gamma=90^\circ$ and a $\cos\theta_{max}=0.98$.}
\end{figure*}

In Fig.~\ref{Energy} we present a study of the phase behavior of the two colour solution, C2(1), since it has the least colours and still assembles all structures. For simplicity, we restrict ourselves to the parameters $\cos\theta_{max}=0.98$ and $\gamma=90^\circ$, a combination which favors the snub cube structure. In panel \emph{a} we plot the Energy-vs-density, $E(\rho)$, curve at different temperatures, where we observe a non-monotonic behavior of the average energy $E(\rho)$ with increasing density. This behaviour is characteristic of the self-assembly of finite-size aggregates~\cite{Sciortino2009}: at low densities the system is in a gas phase of mostly monomers (unbounded particles); with increasing density particles start to aggregate and the energy decreases approaching the value $E=-5/2$ (in units of $\epsilon$) which corresponds to an ideal gas phase of fully formed aggregates; for larger densities the gas phase competes with a percolated liquid phase that, due to geometric constraints originating from the patches arrangement on the surface of the particle, cannot form all available bonds and has thus higher energy than the gas phase. A thermodynamic motivation for the link between a $E(\rho)$ minimum and phase separation is discussed in Ref.~\cite{Russo2021}. Fig.~\ref{Energy}b shows shapshots of Monte Carlo simulations of increasing densities and at three different temperatures, displaying the transition between a gas of monomers, to a gas of snub cubes, to a percolated liquid phase.

In Fig.~\ref{Energy}c we plot the fraction of monomers as a functions of density for the same state points as above, and additionally for the designs C4(1) and C5 at $T=0.1$. From this, we measure the critical micelle concentration (CMC), defined as the number density at $50\%$ of particles are in a monomeric state ($\rho_1=0.5$). The CMC is plotted with (green) squares in Fig.~\ref{Energy}d, where it is seen to have an exponential behaviour as a function of the inverse temperature, $1/T$. We also plot the points corresponding to the minima of the energy, which also show a similar exponential increase. The slope of these curves is well captured by 
the mean-field prediction (blue symbols)~\cite{Kraft2012}

\begin{equation}\label{meanfield}
\rho_1=\frac{1}{V_b}\exp[\frac{c(n)}{(n-1)} \frac{\varepsilon}{k_B T}]
\end{equation}

\noindent where $V_b$ is the bonding volume for the Kern-Frenkel interaction and $c(n)/(n-1)$ is the average number of bonds per particle in the clusters. Using $c(n)/(n-1)$ as a fitting parameter~\cite{Hatch2016}, we employ a least-squares minimization to fit the simulation results for the CMC. We find a value of $c(n)/(n-1)\approx 1.728\pm 0.006$ best approximates the line shown in Fig.~\ref{Energy}.

\subsection{Yields of designs with multiple species}

\begin{figure*}[!t]
	\includegraphics{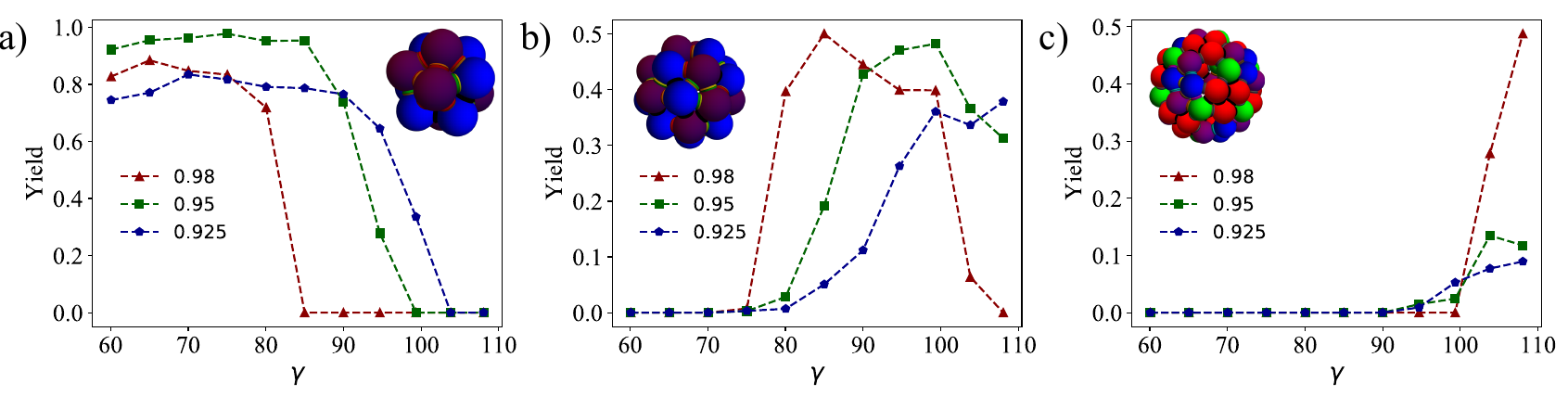}
	\caption{\label{yieldSM} Average yield of the icosahedron, snub cube and snub dodecahedron (in Fig.~1) as a function of $\gamma$ (a, b and c plots respectively). These yields correspond to the results shown in Fig~5 for the designs with multiple species and mutual exclusions.}
\end{figure*}

In Fig.~\ref{yieldSM} we show the yields measured for the designs with multiple species presented in subsection C of the SAT designs section. We show three different yield curves for the three different shells. Each curve corresponds to a different value of $\cos\theta_{max}$.

\pagebreak

\pagebreak

\begin{table}[!t]
	\begin{center}
		\begin{tabular}{ cccc } 
			\hline
			Orientation $o$ & Mapping $\phi_o$ \\
			\hline
			1 & (1,2,3,4,5) \\ 
			2 & (5,1,2,3,4) \\ 
			3 & (4,5,1,2,3) \\ 
			4 & (3,4,5,1,2) \\
			5 & (2,3,4,5,1) \\
			\hline
		\end{tabular}
		\caption{Mapping of the orientation to the patche numbers for the icosahedron.}
	\end{center}
\end{table}

\pagebreak

\begin{table}
	\begin{tabular}{ cccc } 
		\hline
		Position $l_i$ & Patch $s_i$ & Position $l_j$ & Patch $s_j$ \\
		\hline
		1 & 1 & 3 & 3 \\
		1 & 2 & 9 & 2 \\
		1 & 3 & 5 & 3 \\
		1 & 4 & 6 & 1 \\
		1 & 5 & 10 & 2 \\
		2 & 1 & 4 & 3 \\
		2 & 2 & 11 & 1 \\
		2 & 3 & 7 & 3 \\
		2 & 4 & 8 & 1 \\
		2 & 5 & 12 & 3 \\
		3 & 1 & 7 & 1 \\
		3 & 2 & 9 & 3 \\
		3 & 4 & 10 & 1 \\
		3 & 5 & 8 & 3 \\
		4 & 1 & 5 & 1 \\
		4 & 2 & 11 & 2 \\
		4 & 4 & 12 & 2 \\
		4 & 5 & 6 & 3 \\
		5 & 2 & 6 & 2 \\
		5 & 4 & 9 & 1 \\
		5 & 5 & 11 & 3 \\
		6 & 4 & 12 & 1 \\
		6 & 5 & 10 & 3 \\
		7 & 2 & 8 & 2 \\
		7 & 4 & 11 & 5 \\
		7 & 5 & 9 & 4 \\
		8 & 4 & 10 & 5 \\
		8 & 5 & 12 & 4 \\
		9 & 5 & 11 & 4 \\
		10 & 4 & 12 & 5 \\
		\hline
	\end{tabular}
	\caption{Topology of the icosahedron.}
\end{table}

\pagebreak
\FloatBarrier

\begin{table}
	\begin{tabular}{ cccc } 
		\hline
		Position $l_i$ & Patch $s_i$ & Position $l_j$ & Patch $s_j$ \\
		\hline
		1 & 1 & 14 & 1 \\
		1 & 2 & 24 & 3 \\
		1 & 3 & 20 & 2 \\
		1 & 4 & 5 & 5 \\
		1 & 5 & 6 & 4 \\
		2 & 1 & 13 & 1 \\
		2 & 2 & 21 & 3 \\
		2 & 3 & 17 & 2 \\
		2 & 4 & 6 & 5 \\
		2 & 5 & 5 & 4 \\
		3 & 1 & 16 & 1 \\
		3 & 2 & 23 & 3 \\
		3 & 3 & 19 & 2 \\
		3 & 4 & 7 & 5 \\
		3 & 5 & 8 & 4 \\
		4 & 1 & 15 & 1 \\
		4 & 2 & 22 & 3 \\
		4 & 3 & 18 & 2 \\
		4 & 4 & 8 & 5 \\
		4 & 5 & 7 & 4 \\
		5 & 1 & 20 & 1 \\
		5 & 2 & 9 & 3 \\
		5 & 3 & 13 & 2 \\
		6 & 1 & 17 & 1 \\
		6 & 2 & 10 & 3 \\
		6 & 3 & 14 & 2 \\
		7 & 1 & 19 & 1 \\
		7 & 2 & 11 & 3 \\
		7 & 3 & 15 & 2 \\
		8 & 1 & 18 & 1 \\
		8 & 2 & 12 & 3 \\
		8 & 3 & 16 & 2 \\
		9 & 1 & 23 & 1 \\
		9 & 2 & 13 & 3 \\
		9 & 4 & 20 & 5 \\
		9 & 5 & 19 & 4 \\
		10 & 1 & 22 & 1 \\
		10 & 2 & 14 & 3 \\
		10 & 4 & 17 & 5 \\
		10 & 5 & 18 & 4 \\
		11 & 1 & 24 & 1 \\
		11 & 2 & 15 & 3 \\
		11 & 4 & 19 & 5 \\
		11 & 5 & 20 & 4 \\
		12 & 1 & 21 & 1 \\
		12 & 2 & 16 & 3 \\
		12 & 4 & 18 & 5 \\
		12 & 5 & 17 & 4 \\
		13 & 4 & 23 & 5 \\
		13 & 5 & 21 & 4 \\
		14 & 4 & 22 & 5 \\
		14 & 5 & 24 & 4 \\
		15 & 4 & 24 & 5 \\
		15 & 5 & 22 & 4 \\
		16 & 4 & 21 & 5 \\
		16 & 5 & 23 & 4 \\
		17 & 3 & 21 & 2 \\
		18 & 3 & 22 & 2 \\
		19 & 3 & 23 & 2 \\
		20 & 3 & 24 & 2 \\
		\hline
	\end{tabular}
	\caption{Topology of the snub cube.}
\end{table}

\pagebreak
\FloatBarrier

\begin{longtable}{ cccc }
	\hline
	Position $l_i$ & Patch $s_i$ & Position $l_j$ & Patch $s_j$ \\
	\hline
	1 & 1 & 9 & 2 \\
	1 & 2 & 25 & 1 \\
	1 & 3 & 19 & 3 \\
	1 & 4 & 27 & 5 \\
	1 & 5 & 5 & 4 \\
	2 & 1 & 10 & 2 \\
	2 & 2 & 26 & 1 \\
	2 & 3 & 20 & 3 \\
	2 & 4 & 28 & 5 \\
	2 & 5 & 6 & 4 \\
	3 & 1 & 5 & 2 \\
	3 & 2 & 29 & 1 \\
	3 & 3 & 13 & 3 \\
	3 & 4 & 43 & 5 \\
	3 & 5 & 9 & 4 \\
	4 & 1 & 6 & 2 \\
	4 & 2 & 30 & 1 \\
	4 & 3 & 14 & 3 \\
	4 & 4 & 44 & 5 \\
	4 & 5 & 10 & 4 \\
	5 & 1 & 29 & 2 \\
	5 & 3 & 9 & 3 \\
	5 & 5 & 15 & 4 \\
	6 & 1 & 30 & 2 \\
	6 & 3 & 10 & 3 \\
	6 & 5 & 16 & 4 \\
	7 & 1 & 15 & 2 \\
	7 & 2 & 45 & 1 \\
	7 & 3 & 11 & 3 \\
	7 & 4 & 49 & 5 \\
	7 & 5 & 29 & 4 \\
	8 & 1 & 16 & 2 \\
	8 & 2 & 46 & 1 \\
	8 & 3 & 12 & 3 \\
	8 & 4 & 50 & 4 \\
	8 & 5 & 30 & 4 \\
	9 & 1 & 25 & 2 \\
	9 & 5 & 17 & 4 \\
	10 & 1 & 26 & 2 \\
	10 & 5 & 18 & 4 \\
	11 & 1 & 52 & 2 \\
	11 & 2 & 49 & 1 \\
	11 & 4 & 45 & 5 \\
	11 & 5 & 54 & 4 \\
	12 & 1 & 51 & 2 \\
	12 & 2 & 50 & 3 \\
	12 & 4 & 46 & 5 \\
	12 & 5 & 53 & 4 \\
	13 & 1 & 53 & 2 \\
	13 & 2 & 43 & 1 \\
	13 & 4 & 29 & 5 \\
	13 & 5 & 51 & 4 \\
	14 & 1 & 54 & 2 \\
	14 & 2 & 44 & 1 \\
	14 & 4 & 30 & 5 \\
	14 & 5 & 52 & 4 \\
	15 & 1 & 45 & 2 \\
	15 & 3 & 29 & 3 \\
	15 & 5 & 39 & 4 \\
	16 & 1 & 46 & 2 \\
	16 & 3 & 30 & 3 \\
	16 & 5 & 40 & 4 \\
	17 & 1 & 59 & 2 \\
	17 & 2 & 31 & 1 \\
	17 & 3 & 25 & 3 \\
	17 & 5 & 55 & 4 \\
	18 & 1 & 60 & 2 \\
	18 & 2 & 32 & 1 \\
	18 & 3 & 26 & 3 \\
	18 & 5 & 56 & 4 \\
	19 & 1 & 37 & 2 \\
	19 & 2 & 27 & 1 \\
	19 & 4 & 25 & 5 \\
	19 & 5 & 35 & 4 \\
	20 & 1 & 38 & 2 \\
	20 & 2 & 28 & 1 \\
	20 & 4 & 26 & 5 \\
	20 & 5 & 36 & 4 \\
	21 & 1 & 39 & 2 \\
	21 & 2 & 41 & 1 \\
	21 & 3 & 23 & 3 \\
	21 & 4 & 47 & 5 \\
	21 & 5 & 45 & 4 \\
	22 & 1 & 40 & 2 \\
	22 & 2 & 42 & 1 \\
	22 & 3 & 24 & 3 \\
	22 & 4 & 48 & 5 \\
	22 & 5 & 46 & 4 \\
	23 & 1 & 56 & 2 \\
	23 & 2 & 47 & 1 \\
	23 & 4 & 41 & 5 \\
	23 & 5 & 60 & 4 \\
	24 & 1 & 55 & 2 \\
	24 & 2 & 48 & 1 \\
	24 & 4 & 42 & 5 \\
	24 & 5 & 59 & 4 \\
	25 & 4 & 31 & 5 \\
	26 & 4 & 32 & 5 \\
	27 & 2 & 37 & 1 \\
	27 & 3 & 41 & 3 \\
	27 & 4 & 39 & 5 \\
	28 & 2 & 38 & 1 \\
	28 & 3 & 42 & 3 \\
	28 & 4 & 40 & 5 \\
	31 & 2 & 59 & 1 \\
	31 & 3 & 34 & 3 \\
	31 & 4 & 57 & 5 \\
	32 & 2 & 60 & 1 \\
	32 & 3 & 33 & 3 \\
	32 & 4 & 58 & 5 \\
	33 & 1 & 35 & 2 \\
	33 & 2 & 58 & 1 \\
	33 & 4 & 60 & 5 \\
	33 & 5 & 37 & 4 \\
	34 & 1 & 36 & 2 \\
	34 & 2 & 57 & 1 \\
	34 & 4 & 59 & 5 \\
	34 & 5 & 38 & 4 \\
	35 & 1 & 58 & 2 \\
	35 & 3 & 37 & 3 \\
	35 & 5 & 57 & 4 \\
	36 & 1 & 57 & 2 \\
	36 & 3 & 38 & 3 \\
	36 & 5 & 58 & 4 \\
	37 & 5 & 41 & 4 \\
	38 & 5 & 42 & 4 \\
	39 & 1 & 41 & 2 \\
	39 & 3 & 45 & 3 \\
	40 & 1 & 42 & 2 \\
	40 & 3 & 46 & 3 \\
	43 & 2 & 53 & 1 \\
	43 & 3 & 48 & 3 \\
	43 & 4 & 55 & 5 \\
	44 & 2 & 54 & 1 \\
	44 & 3 & 47 & 3 \\
	44 & 4 & 56 & 5 \\
	47 & 2 & 56 & 1 \\
	47 & 4 & 54 & 5 \\
	48 & 2 & 55 & 1 \\
	48 & 4 & 53 & 5 \\
	49 & 2 & 52 & 1 \\
	49 & 3 & 50 & 1 \\
	49 & 4 & 51 & 5 \\
	50 & 2 & 51 & 1 \\
	50 & 5 & 52 & 5 \\
	51 & 3 & 53 & 3 \\
	52 & 3 & 54 & 3 \\
	55 & 3 & 59 & 3 \\
	56 & 3 & 60 & 3 \\
	57 & 3 & 58 & 3 \\
	\hline
	\caption{Topology of the snub dodecahedron.}
\end{longtable}

\pagebreak
\FloatBarrier

\begin{table}
	\begin{center}
		\begin{tabular}{ cccc } 
			\hline
			& N2(1) \\
			\hline
			Patch number & Colour & Interaction \\
			\hline
			1 & A & (A,A) \\ 
			2 & A & (A,A) \\ 
			3 & A & (A,A) \\ 
			4 & B & (B,B) \\
			5 & B & (B,B) \\
			\hline
			\hline
			& N2(2) \\
			\hline
			Patch number & Colour & Interaction \\
			\hline
			1 & A & (A,A) \\ 
			2 & B & (B,B) \\ 
			3 & B & (B,B) \\ 
			4 & A & (A,A) \\
			5 & A & (A,A) \\
			\hline
			\hline
			& N3(1) \\
			\hline
			Patch number & Colour & Interaction \\
			\hline
			1 & A & (A,A) \\ 
			2 & A & (A,A) \\ 
			3 & A & (A,A) \\ 
			4 & B & (B,C) \\
			5 & C & (C,B) \\
			\hline
			\hline
			& N3(2) \\
			\hline
			Patch number & Colour & Interaction \\
			\hline
			1 & B & (B,B) \\ 
			2 & A & (A,A) \\ 
			3 & A & (A,A) \\ 
			4 & C & (C,C) \\
			5 & C & (C,C) \\
			\hline
			\hline
			& N4(1) \\
			\hline
			Patch number & Colour & Interaction \\
			\hline
			1 & B & (B,B) \\ 
			2 & A & (A,D) \\ 
			3 & D & (D,A) \\ 
			4 & C & (C,C) \\
			5 & C & (C,C) \\
			\hline
			\hline
			& N4(2) \\
			\hline
			Patch number & Colour & Interaction \\
			\hline
			1 & B & (B,B) \\ 
			2 & C & (C,C) \\ 
			3 & C & (C,C) \\ 
			4 & A & (A,D) \\
			5 & D & (D,A) \\
			\hline
			\hline
			& N5 \\
			\hline
			Patch number & Colour & Interaction \\
			\hline
			1 & A & (A,A) \\ 
			2 & C & (C,B) \\ 
			3 & B & (B,C) \\ 
			4 & D & (D,E) \\
			5 & E & (E,D) \\
			\hline
		\end{tabular}
		\caption{Different solutions used in Fig.~4. The first column indicates the patch number, the second the colour associated with it, and the third column indicates the corresponding bond formed, the first letter corresponds to the colour of the patch and the second number to the colour it interacts with.}
	\end{center}
\end{table}

\pagebreak
\FloatBarrier

\begin{table}
	\begin{center}
		\begin{tabular}{ cccc } 
			\hline
			& Icosahedron \\
			\hline
			Patch number & Colour & Interaction \\
			\hline
			& Specie 1 \\
			\hline
			1 & C & (C,B) \\ 
			2 & F & (F,F) \\ 
			3 & C & (C,B) \\ 
			4 & F & (F,F) \\
			5 & D & (D,A) \\
			\hline
			& Specie 2 \\
			\hline
			1 & C & (C,B) \\ 
			2 & B & (B,C) \\ 
			3 & A & (A,D) \\ 
			4 & B & (B,C) \\
			5 & B & (B,C) \\
			\hline
			\hline
			& Snub cube \\
			\hline
			Patch number & Colour & Interaction \\
			\hline
			& Specie 1 \\
			\hline
			1 & F & (F,F) \\ 
			2 & D & (D,A) \\ 
			3 & A & (A,D) \\ 
			4 & F & (F,F) \\
			5 & C & (C,B) \\
			\hline
			& Specie 2 \\
			\hline
			1 & F & (F,F) \\ 
			2 & D & (D,A) \\ 
			3 & A & (A,D) \\ 
			4 & B & (B,C) \\
			5 & F & (F,F) \\
			\hline
			\hline
			& Snub dodecahedron \\
			\hline
			Patch number & Colour & Interaction \\
			\hline
			& Specie 1 \\
			\hline
			1 & K & (K,K) \\ 
			2 & L & (L,E) \\ 
			3 & D & (D,D) \\ 
			4 & J & (J,I) \\
			5 & H & (H,G) \\
			\hline
			& Specie 2 \\
			\hline
			1 & F & (F,A) \\ 
			2 & A & (A,F) \\ 
			3 & F & (F,A) \\ 
			4 & G & (G,H) \\
			5 & G & (G,H) \\
			\hline
			& Specie 3 \\
			\hline
			1 & B & (B,C) \\ 
			2 & K & (K,K) \\ 
			3 & A & (A,F) \\ 
			4 & H & (H,G) \\
			5 & H & (H,G) \\
			\hline
			& Specie 4 \\
			\hline
			1 & E & (E,L) \\ 
			2 & C & (C,B) \\ 
			3 & A & (A,F) \\ 
			4 & H & (H,G) \\
			5 & I & (I,J) \\
			\hline
		\end{tabular}
		\caption{Different solutions used in Fig~5. The first column indicates the patch number, the second the colour associated with it, and the third column indicates the corresponding bond formed, the first number corresponds to the colour of the patch and the second number to the colour it interacts with.}
	\end{center}
\end{table}

\end{document}